\documentclass[useAMS,usegraphicx]{mn2e}
\usepackage{rotate}
\usepackage{rotating}
\usepackage{times}
\newif\ifAMStwofonts
\AMStwofontstrue

\def\gs{\mathrel{\hbox{\rlap{\hbox{\lower4pt\hbox{$\sim$}}}\hbox{$>$}}}}
\def\ls{\mathrel{\hbox{\rlap{\hbox{\lower4pt\hbox{$\sim$}}}\hbox{$<$}}}}

\def\suzaku{{\it Suzaku}}

\def\rxte{{\it RXTE}}
\def\xmm{{\it XMM-Newton}}

\def\et{{et al.\ }}

\def\mrk79{{Mrk~79}}
\def\ugc{{UGC~3973}}

\def\3c{{3C~273}}
\def\rg{{\thinspace r_{\rm g}}}
\def\risco{{\thinspace r_{\rm ISCO}}}
\def\fvar{{F_{\rm var}}}
\def\chidof{{\chi^2_\nu/{\rm dof}}}
\def\redchi{{\chi^2_\nu}}
\def\delchi{{\Delta\chi^2}}
\def\feka{{Fe~K$\alpha$}}
\def\fela{{Fe~L$\alpha$}}

\def\fekb{{Fe~K$\beta$}}
\def\ovii{{O~\textsc{vii}}}
\def\oviii{{O~\textsc{viii}}}

\def\fexxv{{Fe~\textsc{xxv}}}
\def\fexxvi{{Fe~\textsc{xxvi}}}

\def\deg{^{\circ}}
\def\A{{\rm\thinspace \AA}}
\def\cm{{\rm\thinspace cm}}
\def\erg{{\rm\thinspace erg}}
\def\eV{{\rm\thinspace eV}}

\def\ga{{\rm\thinspace gauss}}

\def\K{{\rm\thinspace K}}
\def\keV{{\rm\thinspace keV}}
\def\km{{\rm\thinspace km}}

\def\Msun{\hbox{$\rm\thinspace M_{\odot}$}}

\def\pc{{\rm\thinspace pc}}
\def\s{{\rm\thinspace s}}
\def\ks{{\rm\thinspace ks}}
\def\ps{{\rm\thinspace s^{-1}}}

\def\cmps{\hbox{$\cm\s^{-1}\,$}}

\def\ergpscmps{\hbox{$\erg\cm^{-2}\s^{-1}\,$}}

\def\ergps{\hbox{$\erg\s^{-1}\,$}}

\def\kmps{\hbox{$\km\ps\,$}}

\def\pscm{\hbox{$\cm^{-2}\,$}}

\def\pccm{\hbox{$\cm^{-3}\,$}}

\title[Multi-epoch X-ray observations of \mrk79]
      {
Multi-epoch X-ray observations of the Seyfert~1.2 galaxy \mrk79: 
bulk motion of the illuminating X-ray source 
      }

\author[L. C. Gallo et al.]
       {L. C. Gallo,$^1$ 
	G. Miniutti,$^2$  
	J. M. Miller,$^3$  
        L. W. Brenneman,$^4$
	A. C. Fabian,$^5$ 
        M. Guainazzi,$^6$
\newauthor
        and C. S. Reynolds$^7$
        \\ 
$^{1}$ Department of Astronomy and Physics, Saint Mary's University, 923 Robie Street, Halifax, NS, B3H 3C3, Canada \\
$^{2}$ Centro de Astrobiologia (CSIC-INTA), Dep. de Astrofisica; LAEFF, PO Box 78, E-28691, Villanueva de la Ca\~nada, Madrid, Spain \\
$^{3}$ Department of Astronomy, University of Michigan, 500 Church Street, Ann Arbor, MI 48109, USA \\
$^{4}$ Harvard-Smithsonian Center for Astrophysics, MS 67, 60 Garden Street, Cambridge, MA, 01238, USA \\
$^{5}$ Institute of Astronomy, University of Cambridge, Madingley Road, Cambridge CB3 0HA\\
$^{6}$ European Space Astronomy Centre of ESA, Apartado 50727, E-28080 Madrid, Spain \\
$^{7}$ Department of Astronomy and the Maryland Astronomy Center for Theory and Computation, University of Maryland, College Park, MD 20742, USA \\
}
\date{Accepted. Received. }
\pagerange{\pageref{firstpage}--\pageref{lastpage}}
\pubyear{2010}
\begin{document}
\maketitle
\label{firstpage}

\begin{abstract}
Multi-epoch X-ray spectroscopy ($0.3-25\keV$) of the Seyfert~1.2 galaxy \mrk79\ (\ugc) spanning nearly
eight years and a factor of three in broadband flux are analysed.  The data are obtained at
seven epochs with either \xmm\ or \suzaku.  Comparison with contemporaneous \rxte\ monitoring
indicate that all flux states of \mrk79\ are represented by the data.  The spectra are fitted in a 
self-consistent manner adopting a power law and ionised reflection to describe the broadband 
continuum.  Modification of the spectra by a distant photoionised medium, seen predominantly in emission,
are also included.  
Under the assumption that the inner disk is at the innermost stable
circular orbit, our blurred reflection models give a spin of $a = 0.7 \pm
0.1$.  The reflection component in each spectrum is weaker than predicted
by simple reflection models.  If the illuminating X-ray emission is
produced by flares above the disk that move at mildly relativistic
velocities, however, diminished reflection is expected.  Light bending due
to strong gravity near black holes can influence how the
illuminating and reflected flux are observed; variations in \mrk79\ do not
suggest that light bending is important in this source.
\end{abstract}



\section{Introduction}
\label{sect:intro}

The X-ray spectra of type I active galactic nuclei (AGNs) can exhibit significant 
variability from one epoch to the next and the general spectral shape can normally be 
associated with the flux state of the AGN.  For example, in the low-flux state the 
power law emission above $\sim 2\keV$ is flatter and the excess above the power law 
at energies $\ls2\keV$ (i.e. the so called soft-excess) is relatively stronger than 
in the high-flux state.
In addition, the $2-10\keV$ band often displays complexity in the form
of spectral curvature or spectral drops in the low-flux state (e.g. Gallo \et 2004; Grupe \et 2008;
Bachev \et 2009).  However,
defining the flux ``state'' for AGN is more complicated than for black hole binaries 
because the characteristic time scales are much longer in AGN.  Long-term
monitoring (i.e. years) is generally required to properly determine flux states.  Gallo (2006) even suggested
that the hypothetical UV-to-X-ray slope ($\alpha_{ox}$) may be better suited for defining 
the X-ray low state rather than X-ray flux alone.  Consequently, it 
is challenging to 
gain understanding of a particular system from single-epoch observation without information 
of its typical variability behaviour.

An advantage of multi-epoch modelling is that model parameters that vary on
different characteristic time scales can be examined and constrained accordingly.   
While black hole spin, disc inclination and elemental abundances are not expected to 
change; power law spectral slopes can vary on day-long scales (perhaps more
rapidly) (e.g. Sobolewska \& Papadakis 2009; Iwasawa \et 2010); 
and the inner disc ionisation parameter and the absorber covering fraction
have been suggested to vary rapidly ($10^{3-4}\s$) (Ponti \et 2006, 2010; Risaliti \et 2009).

In this work multi-epoch X-ray observations of the Seyfert~1.2 galaxy \mrk79\ 
(\ugc; $z=0.022189$) are examined to study the nature of its X-ray emission and variability.  
\mrk79\ is well-studied in the optical, benefiting
from efforts to measure its black hole mass by reverberation mapping (e.g. Peterson \et
2004).
However, despite being relatively  bright ($\gs 10^{-11}\ergpscmps$ between
$0.5-10\keV$) \mrk79\ is poorly studied in the X-rays.  Snap-shot observations of \mrk79\
were done in 2000 and 2001 with \xmm, marking the first time the spectrum above $2\keV$ was
analysed (Gallo \et 2005; 2006). The observations revealed
a soft-excess, possible warm absorption, and a slightly broadened \feka\ emission line. 
Gallo \et (2005) report a possible $8\keV$ emission feature in the 2001 observation, 
but this is not confirmed with the deeper observations presented here.  

In the following section the observations and data reduction
are described.  In Section~3, the data sets are examined and characterised. 
The RGS spectrum is analysed in Section~4 to understand the nature of the
ionised plasma.  The high-energy spectrum above $3\keV$ is examined in Section~5.
A self-consistent model simultaneously describing all the data is presented in Section~6.
We discuss our results and conclusion in Section~7 and 8, respectively.

\section{Observations and data reduction}
\label{sect:data}
\mrk79\ was observed with \xmm\ (Jansen \et 2001) on five occasions between
2000 and 2008, and once in 2007 with \suzaku\ (Mitsuda \et 2007).  A summary of the 
observations is provided in Table~\ref{tab:obslog} along with the nomenclature
we adopt throughout this work.  

The EPIC pn (Str\"uder \et 2001) and MOS (MOS1 and MOS2;
Turner \et 2001) cameras were operated in small-window mode
with the medium filter in place for all the observations except XMM6 when the
thin filter was used.
The Reflection Grating Spectrometers (RGS1 and RGS2; den Herder \et 2001)
also collected data during this time, as did the Optical Monitor
(OM; Mason \et 2001).

The \xmm\ Observation Data Files (ODFs) from all observations
were processed to produce calibrated event lists using the \xmm\ 
Science Analysis System ({\tt SAS v9.0.0}).
Unwanted hot, dead, or flickering pixels were removed as were events due to
electronic noise.  Event energies were corrected for charge-transfer
inefficiencies.  EPIC response matrices were generated using the {\tt SAS}
tasks {\tt ARFGEN} and {\tt RMFGEN}.  Light curves were extracted from these
event lists to search for periods of high background flaring.
Significant flaring was detected during the final $\sim18\ks$ of XMM6 and data during these
periods have been neglected.
The total amount of good pn exposure is listed in Table~\ref{tab:obslog}.
Source photons were extracted from a circular region 35$^{\prime\prime}$ across
and centred on the source.
The background photons were extracted from an off-source region on the same CCD.
Single and double events were selected for the pn detector, and
single-quadruple events were selected for the MOS.
The MOS and pn data at each epoch are compared for consistency and determined to
be in agreement within known uncertainties (Guainazzi \et 2010).  
For simplicity in analysis and presentation only the pn data are discussed here.

The RGS spectra were extracted using the {\tt SAS} task {\tt RGSPROC} and 
response matrices were generated using {\tt RGSRMFGEN}.
The OM operated in imaging mode with various filters in place at different epochs.
However, an ultraviolet image was obtained at every epoch in either the $UVW1$ ($2450-3200\A$) 
or $UVW2$ ($1800-2250\A$) filter.

\suzaku\ observed \mrk79\ on 2007 April 03.  The two front-illuminated (FI) CCDs
(XIS0 and XIS3), the back-illuminated (BI) CCD (XIS1), and the HXD-PIN all functioned
normally and collected data during this period.  The target was observed in the
XIS-nominal position.

Cleaned event files from version 2 processed data were used in the analysis and data
products were extracted using {\tt xselect}.
For each XIS chip, source counts were extracted from a $4^{\prime}$ circular region centred on 
the target.  Background counts were taken from surrounding regions on the chip.  Response files
(rmf and arf) were 
generated using {\tt xisrmfgen} and {\tt xissimarfgen}.  
A spectrum of the $^{55}$Fe calibration source was extracted from each CCD.  The Mn~K$\alpha$
line was fit with a Gaussian profile to measure the peak energy.  The energies were
within $0.2$ per cent of the expected value ($5.895\keV$) and consistent with current
accuracy of the XIS energy scale.\footnote{http://www.astro.isas.jaxa/suzaku/doc/suzaku\_td/}
After examining for consistency, the 
data from the XIS-FI were combined
to create a single spectrum.

The PIN spectrum was extracted from the HXD data following standard procedures.
A non-X-ray background (NXB) file corresponding to the observation was obtained to generate
good-time interval (GTI) common to the data and NXB.  The data were also corrected for 
detector deadtime.  The resulting PIN exposure was $81\ks$.  The cosmic X-ray background
(CXB) was modelled using the provided flat response files.  The CXB and NXB background files
were combined to create the PIN background spectrum.  In comparison to the source plus
background PIN data, \mrk79\ is detected between $12-25\keV$.  

For the X-ray spectral analysis
the source spectra are grouped such that each bin contains at least 20
counts (typically more). For display purposes data are presented in larger bins.
Spectral fitting was performed using {\tt XSPEC v12.5.0}
(Arnaud 1996).

All parameters are reported in the rest frame of the source unless specified
otherwise.
The quoted errors on the model parameters correspond to a 90\% confidence
level for one interesting parameter (i.e. a $\Delta\chi^2$ = 2.7 criterion).
We note the uncertainties in Section~6 (Table~\ref{tab:fits}) are underestimated.
Given the complexity and slow-fitting nature in the presented model, the uncertainties on 
any one model parameter are calculated while keeping the other variable parameters fixed.
A value for the Galactic column density toward \mrk79\ of
$5.27 \times 10^{20}\pscm$ (Kalberla \et 2005) is adopted in all of the
spectral fits.  
Fluxes for all model components are estimated with the {\tt XSPEC} model
{\tt cflux}.
K-corrected luminosities are calculated using a
Hubble constant of $H_0$=$\rm 70\ km\ s^{-1}\ Mpc^{-1}$ and
a standard flat cosmology with $\Omega_{M}$ = 0.3 and $\Omega_\Lambda$ = 0.7.
The HXD/PIN-XIS cross normalisation is fixed at 1.16.

\begin{table*}
\begin{center}
\caption{\mrk79\ observation log.  Each data set will be referred to in the paper by
the identifier in column 1.  The observatory used to obtain the data and the observation
id are given in column 2 and 3, respectively.  The start date and duration of the observation
are shown in columns 4 and 5, respectively.  Column 6 shows the good time interval for each
observation.  The \suzaku\ exposure is combined for all three XIS.  Column 7 provides the
total source counts in the $0.3-12\keV$ band ($0.5-10\keV$ for \suzaku).
}
\begin{tabular}{ccccccc}                
\hline
(1) & (2) & (3) & (4) & (5) & (6) & (7) \\
Observation   &  Observatory & Observation ID  & Start Date   & Duration  &  Exposure & Counts\\
              &   &                  & (year.mm.dd) &       (s) &       (s) & \\
\hline
XMM1 & \xmm\ & 0103860801 & 2000.10.09 & 6904  & 1680 & 20827 \\
XMM2 & \xmm\ & 0103862101 & 2001.04.26 & 6935  & 3590 & 21454 \\
XMM3 & \xmm\ & 0400070201 & 2006.09.30 & 24000 & 14411 & 238259 \\
XMM4 & \xmm\ & 0400070301 & 2006.11.01 & 21500 & 13993 & 150349 \\
XMM5 & \xmm\ & 0400070401 & 2007.03.19 & 21500 & 13992 & 156980 \\
SUZ1 & \suzaku\ & 702044010  & 2007.04.03 & 160017 & 251127 & 202872 \\ 
     &        &    &            &        & (XIS0+XIS1+XIS3)\\
XMM6 & \xmm\ & 0502091301 & 2008.04.27 & 90400 & 58262 & 136940 \\

\hline
\label{tab:obslog}
\end{tabular}
\end{center}
\end{table*}


\section{Characterisation of the data sets}
\label{sect:char}

\subsection{X-ray flux and spectral variability}
\begin{figure}
\rotatebox{270}
{\scalebox{0.32}{\includegraphics{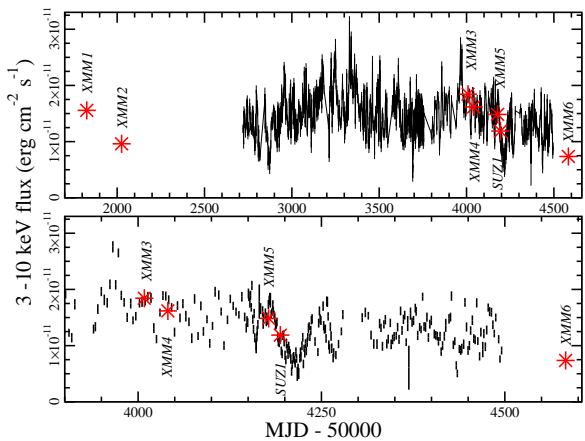}}}
\caption{Upper panel: The $3-10\keV$ long-term \rxte\ light curve is plotted in black.
For comparison the $3-10\keV$ fluxes from each pointed observation are
shown as red stars. Lower panel: The light curve from above is enlarged between
days $3900-4605$ when the pointed observations are more frequent.  The pointed
observations sample all the typical flux states of \mrk79.
}
\label{fig:rxtelc}
\end{figure}

The seven pointed observations were conducted over $\sim7.5$ years and
span a range of $\sim3$ in broadband ($0.5-10\keV$) flux.  Between 2003 and
2008, \mrk79\ underwent monitoring on hourly to daily intervals with \rxte\
(Breedt \et 2009).  The pointed observations are compared with the \rxte\
light curve during this campaign (provided courtesy of P. Uttley) to determine what
flux state is represented by each observation (Figure~\ref{fig:rxtelc}).
XMM2, XMM6, and SUZ1 are representative of low-flux states, with XMM6 comparable to
the lowest states observed with \rxte.  XMM3 is comparable to a high-flux state though
\rxte\ did catch the source about $50$ per cent brighter during occasional flaring
events.  The pointed observations here appear to adequately scrutinise the low-,
high- and intermediate-flux states of \mrk79.

\begin{figure}
\rotatebox{270}
{\scalebox{0.32}{\includegraphics{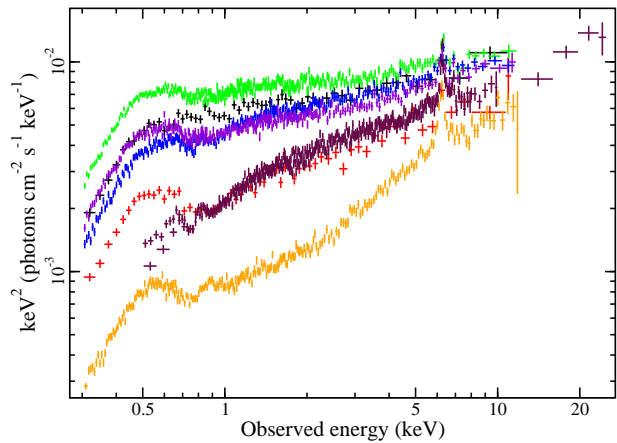}}}
\caption{The spectra from all the pointed observations are shown after 
correcting for the effective area of the detectors.  The spectra become progressively
harder with decreasing flux.  The colour designation is
as follows: XMM1 (black); XMM2 (red); XMM3 (green); XMM4 (blue); XMM5(violet); 
SUZ1 (maroon); and XMM6 (orange).
}
\label{fig:comp}
\end{figure}
The spectra from the pointed observations are compared in Figure~\ref{fig:comp}.
The spectra appear to harden with decreasing flux and a narrow emission feature
at about $6.4\keV$ is seen at all epochs. Though the intermediate-flux states
(XMM1, XMM4, and XMM5) are very comparable in broadband flux, there do appear
to be some slight changes in the spectral shape.

\begin{figure}
\rotatebox{270}
{\scalebox{0.32}{\includegraphics{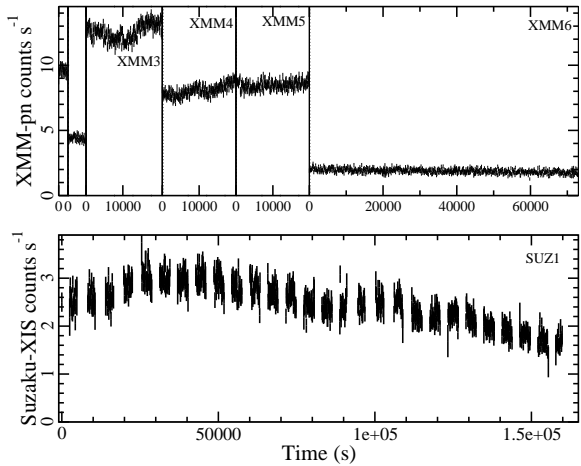}}}
\caption{The short term light curves in $100\s$ bins from all the pointed
observations.  All the \xmm\ pn ($0.2-12\keV$) light curves are presented in the top 
panel.  Zero marks the start time of each light curve.  The count rates
and duration of each curve are on comparable scales.  XMM1 (leftmost panel) and XMM2 
(second panel from left) are short
observations ($<10\ks$).  In the lower panel
the \suzaku\ light curve for the combined XIS in the $0.5-10\keV$ band.  Again, zero marks
the start of the observation.
}
\label{fig:lc}
\end{figure}
The light curves from each observation are shown in Figure~\ref{fig:lc}.
Flux variations on the short time scales (e.g. $\ls1000\s$) probed with \xmm\ are 
present, but are of rather small amplitude.  The $\sim2$--day \suzaku\ observation shows
larger amplitude variations comparable to the fluctuations measured during
the \rxte\ monitoring (Breedt \et 2009).

\begin{figure}
\rotatebox{270}
{\scalebox{0.32}{\includegraphics{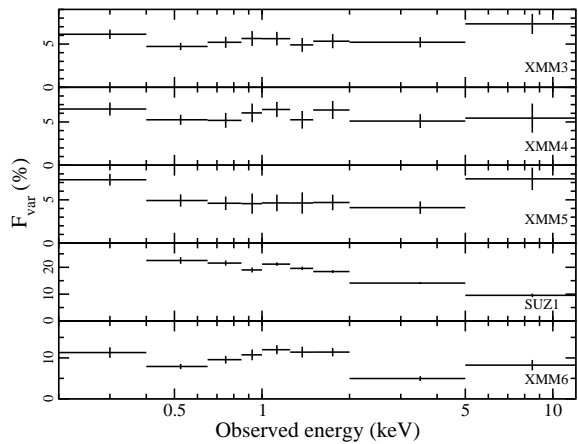}}}
\caption{The normalised rms spectrum calculated for observations of at least
$20\ks$ in duration.  The \xmm\ and \suzaku\ light curves used are in
$750\s$ and $5760\s$ bins, respectively.
}
\label{fig:fvar}
\end{figure}

Spectral variability is tested for by measuring the fractional
variability ($\fvar$) at various energies (e.g. Ponti \et 2004).  For observations
that were at least $20\ks$ in duration the $\fvar$ spectrum is shown in 
Figure~\ref{fig:fvar},  
which confirms the low-amplitude variations
seen in the light curves (Figure~\ref{fig:lc}) and illustrates minimal 
energy-dependent variability during the \xmm\ observations.
Modest spectral variability is observed during the longer \suzaku\ observation
as expected since the variability increases with increasing exposure 
(e.g. Markowitz \& Edelson 2004).
The amplitude of the variations appears to decrease with increasing energy.
Such behavior has been seen in other AGN and attributed to 
various effects (e.g. changing spectral slope or varying the relative
normalisation of the continuum components) (e.g Gallo \et 2007; Ponti \et 2006).  
For our purposes, the
detected spectral variability indicates that caution must be exercised when
fitting the mean \suzaku\ spectrum.

\subsection{Ultraviolet variability}

\begin{figure}
\rotatebox{270}
{\scalebox{0.32}{\includegraphics{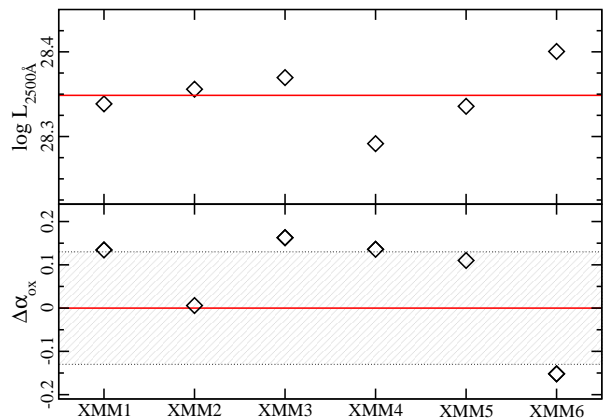}}}
\caption{The $2500\A$ luminosity (upper panel) and $\Delta\alpha_{ox}$
(lower panel) is plotted for each \xmm\ observation.  The red line in the
upper panel identifies the mean UV luminosity.  Based on the photometric
uncertainties only, the error bars on logL$_{2500\A}$ are comparable to the
size of the data points.  
The red line in the lower
panel marks the expected $\alpha_{ox}$ based on the UV luminosity and the
shaded region corresponds to the uncertainties in the Just \et relation. 
Negative values indicate the object is X-ray weak compared to the
UV luminosity.  
}
\label{fig:uv}
\end{figure}

Simultaneous UV photometry with the OM permits examination of the variability at
UV wavelengths.  The average $2500\A$ luminosity over the six \xmm\ observations is
$\sim2.23\times10^{28}\ergps$ with fluctuations on the $\pm10$ per cent
level (Figure~\ref{fig:uv}), significantly lower than seen in the X-rays.

The power law between $2500\A$ and $2\keV$ (i.e. $\alpha_{ox}$) is measured from
the simultaneous UV and X-ray observations.  
According to Just \et (2007), based 
on its UV luminosity the expected UV-X-ray slope for \mrk79\ is 
$\alpha_{ox}(L_{2500\A}) = -1.25\pm0.13$.  
In Figure~\ref{fig:uv}, 
$\Delta\alpha_{ox} = \alpha_{ox} - \alpha_{ox}(L_{2500\A})$ is plotted for each \xmm\
observation to illustrate the deviation of $\alpha_{ox}$ from the expected value for \mrk79
based on its $2500\A$ luminosity at the time.  Most of the time \mrk79\ appears to
be slightly X-ray strong ($\alpha_{ox}(L_{2500\A}) > 0$).  XMM6 is the only observation
in which \mrk79\ is in an X-ray weak state.  According to Gallo (2006), marked
spectral complexity above $2\keV$ (i.e. significant deviations from a power law)
should be present, which is consistent with Figure~\ref{fig:pofit}.
Though the UV luminosity is variable the plunge into an X-ray weak state during
XMM6 is driven by X-ray variability.

\subsection{Phenomenological spectral fits}
\label{sect:ssf}
\begin{figure}
\rotatebox{270}
{\scalebox{0.32}{\includegraphics{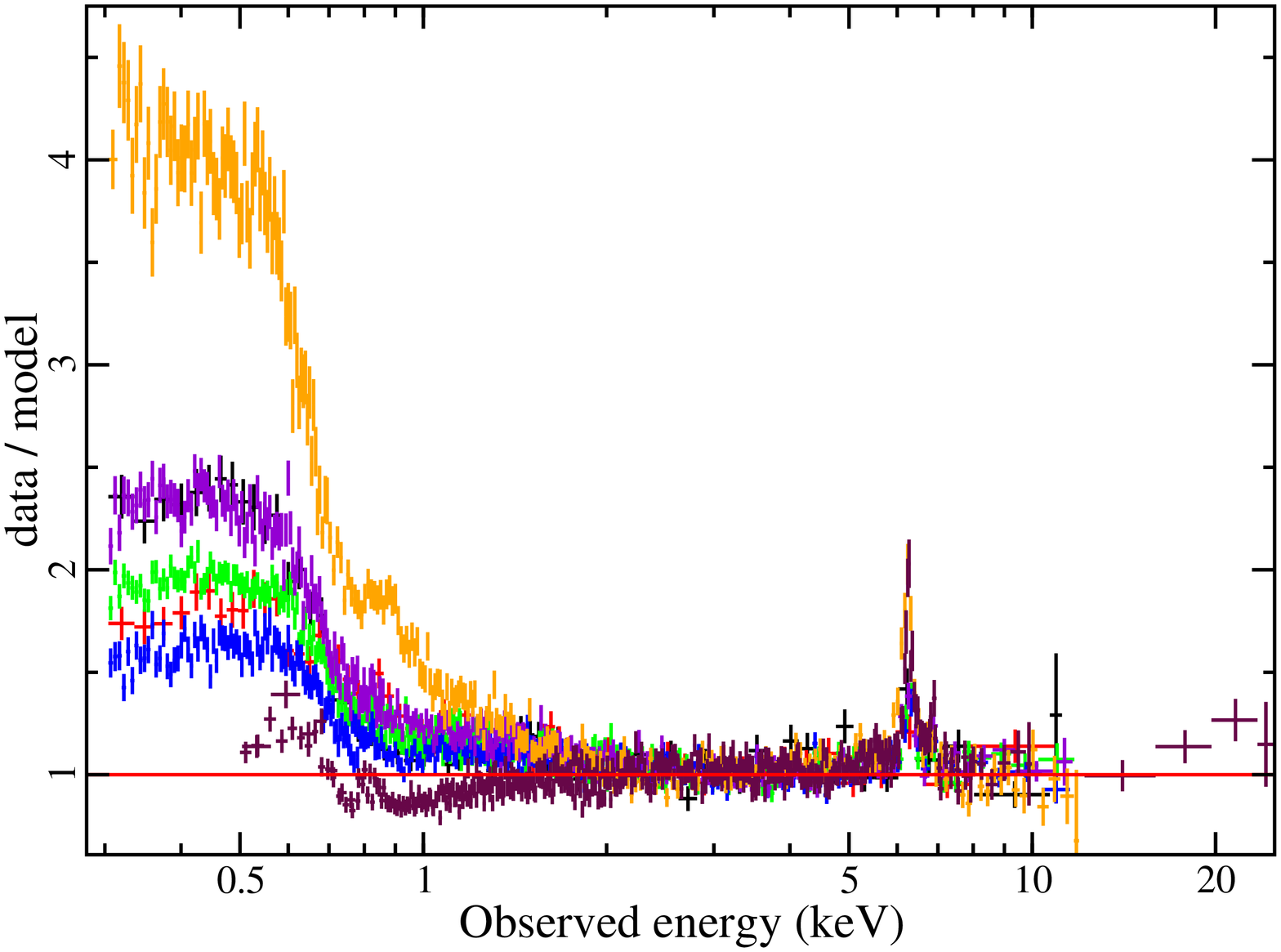}}
\scalebox{0.32}{\includegraphics{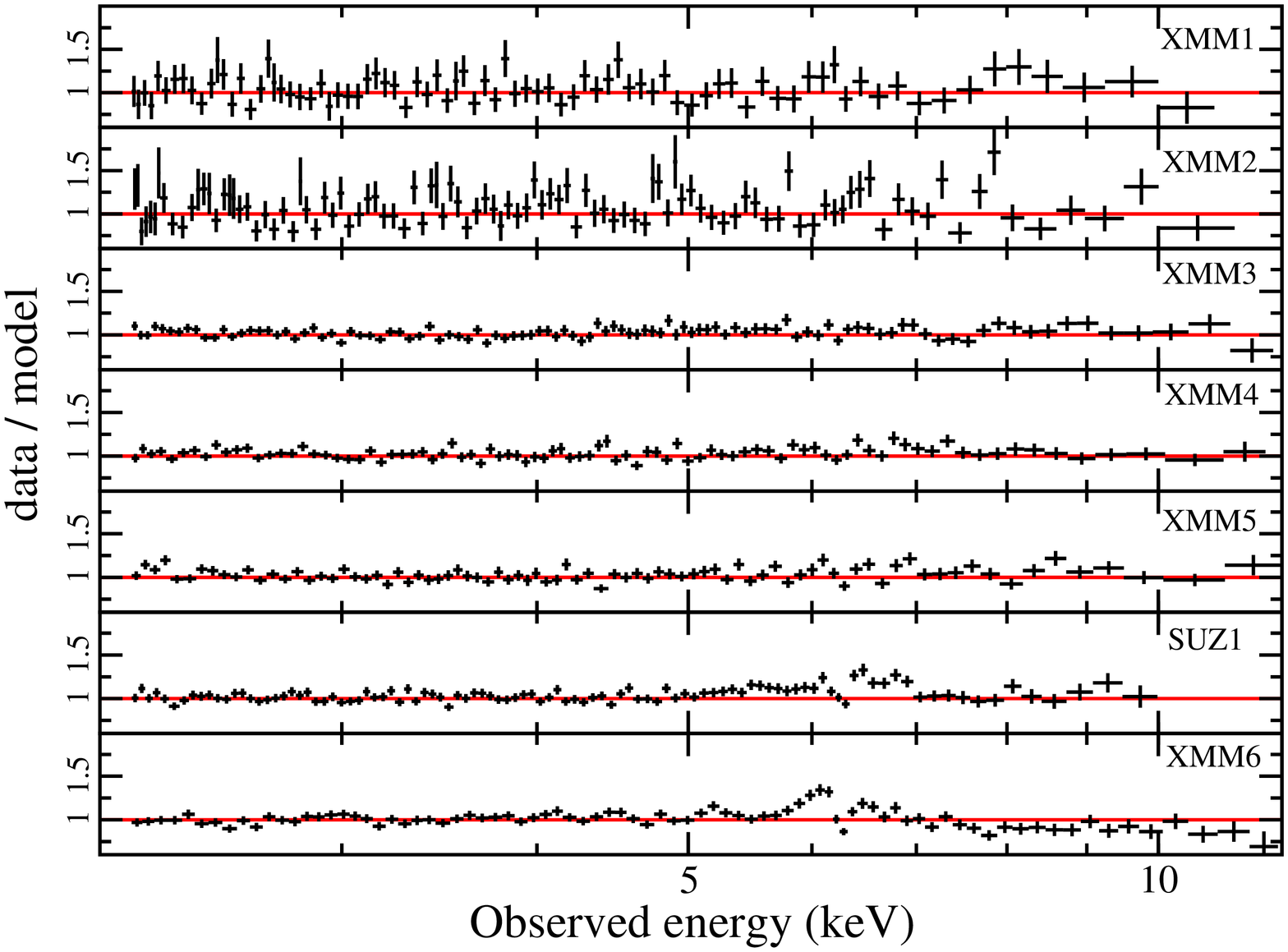}}}
\caption{Upper panel: The residuals (data/model) remaining from fitting each spectrum
with a power law absorbed by Galactic absorption in the $2.2-4.5$ and $7-10\keV$ band, and then
extrapolating over $0.3-25\keV$.  A low-energy excess and narrow emission feature at
$\sim6.4\keV$ are present in all spectra.  The \suzaku\ data show an excess above $12\keV$.
Colours are as described in Figure~\ref{fig:comp}.
Lower panel: The residuals of a power law plus narrow Gaussian profile fit to the $2.2-12\keV$
band. Excess residuals redward of $6.4\keV$ are seen in SUZ1 and XMM6.
}
\label{fig:pofit}
\end{figure}

Fitting a power law to each spectrum in the $2.2-4.5$ and $7-10\keV$ band and extrapolating 
over the $0.3-12\keV$ band ($0.5-25\keV$ for \suzaku) reveals a typical type~I AGN spectrum.
A soft excess is seen below $\sim 2\keV$.  There is a narrow emission feature at $\sim6.4\keV$
as well as a slight excess above $10\keV$ (top panel of Figure~\ref{fig:pofit}). 
Qualitatively, the soft excess and the emission 
feature appear stronger with decreasing flux.  The \suzaku\ data only extend down to
$0.5\keV$ and do not show the full extent of the low-energy excess.

The narrow emission feature at $\sim6.4\keV$ is most likely attributed to \feka\ 
emission from material distant from the black hole.
The spectra above $2.2\keV$ were refit with a power law and narrow Gaussian profile
revealing the residual in the lower panel of Figure~\ref{fig:pofit}.  For XMM1, XMM2,
XMM3, XMM4 and XMM5 the simple fit appears to be acceptable.  For SUZ1 and XMM6
positive residuals remain redward of  $6.4\keV$ that could be interpreted as 
broadened \feka\ emission from the accretion disc.  
For SUZ1, a second Gaussian profile
with a peak energy of $6.46\pm0.14\keV$ ($6.36\pm0.02\keV$ for XMM6) and width of $\sigma=556^{+123}_{-192}\eV$
($142^{+35}_{-18}\eV$ for XMM6) can characterise the residuals. 
We make note of positive residuals 
extending down to $\sim4.5\keV$ in XMM3, but do not consider these as sufficient evidence for
an additional component.

Gallo \et (2006) successfully fit the snap-shot observations of \mrk79\ (XMM1 and XMM2) with
the traditional phenomological model of a black body and power law continuum, with a
high-energy Gaussian profile and low-energy absorption edge.  Applying this model
to the spectrum at each epoch resulted in reasonably good fits ($\redchi=0.89-1.13$).
\begin{figure}
\rotatebox{270}
{\scalebox{0.32}{\includegraphics{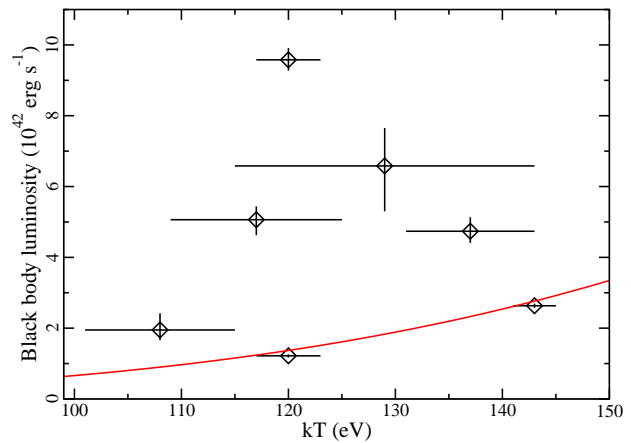}}}
\caption{
The measured black
body luminosity and temperature are plotted for each spectrum.  
The red curve is
the fitted $L\propto T^{4}$ relation (leaving only the normalisation free), 
which clearly does not describe the data.
}
\label{fig:bbpo}
\end{figure}

The soft excess in \mrk79\ can be parameterised as a black body with a 
varying temperature in the range $100-140\eV$, as typically seen in unabsorbed
AGN (e.g. Gierlinski \& Done 2004; Crummy \et 2006).  The measured temperature
is not correlated with the source luminosity in any obvious way and 
does not follow the expected $L \propto T^4$ relation attributed to a standard
accretion disc of constant area (Figure~\ref{fig:bbpo}). The expected 
temperature for a standard accretion disc around \mrk79, assuming a black hole 
mass of $5.24\times10^7\Msun$ (Peterson \et 2004) and $L/L_{Edd}\approx0.09$ 
(following Kaspi \et 2000), peaks at $\sim15\eV$.  Given
the high temperature predicted by the model and apparent variability, a
black body does not provide a physical description for the soft excess in \mrk79.

A narrow Gaussian profile is a rational addition evident from the residuals in
Figure~\ref{fig:pofit}.  The energy is consistent with neutral \feka\ emission and
the flux of the line is comparable within uncertainties at all epochs.  
The average line width corresponds to a {\it FWHM} velocity of $\sim11000\kmps$,
about twice as high as that measured for H$\beta$ from the broad line region 
(BLR; Peterson \et 1998).  This suggests that the dominant component of the
\feka\ emission in \mrk79\ is not coming from a distant torus, but is more
likely emission from the accretion disc and the inner BLR.

The best fit absorption edge energy at most epochs is $\sim720\eV$ ($673\pm6\eV$ for XMM6)
and the depth
of the feature differs at each epoch. 
The energy is lower than expected from \ovii\ ($739\eV$) and not likely
associated with absorption by hot local gas (McKernan \et al. 2004) since the
implied velocities would be higher than the recessional velocity of the AGN.
A more physical treatment is required to interpret this component of the
spectrum (see Section~\ref{sect:multifit}).

\section{The RGS spectrum} 
\label{sect:rgs}

The RGS data were reduced using SASv9.0 \cite{gabriel03} and the
most updated calibration files available at the date the reduction
was performed (October 2009). The data reduction followed standard procedures as
in, e.g. Guainazzi \& Bianchi (2007). The total number of
background subtracted counts in the nominal RGS energy bandpass
(0.3--2~keV) is $674 \pm 19$ and $727 \pm 18$ in the RGS1 and RGS2,
respectively.

Visual inspection of the RGS spectrum does not show the high equivalent width lines
typically observed in the high-resolution spectra of highly obscured
AGN (cf. Kinkhabwala et al. 2002, and references therein). The statistics
of the data does not warrant detailed spectral fitting with physical
models. We have therefore limited ourselves to phenomenological
simultaneous fits of
the combined time-averaged RGS1 and RGS2 spectra.

First we have performed a blind search for absorption lines.
The spectra were fit in 100-channels intervals with a
power-law continuum, photoelectrically absorbed by a column density
held fixed to the value due to intervening gas in our Galaxy.
At the 90 per cent
confidence level for one interesting parameter only one line is detected,
with centroid energy $E_c = 698.2 \pm 1.2\eV$, and intensity
$I = -2.5^{+2.4}_{-2.0} \times 10^{-5}$~cm$^{-1}$~s$^{-2}$.
The 
difference in Cash-statistics between a fit with the continuum only and
the fit after the inclusion of this absorption line is $\Delta C = 26.1$.
The $\Delta C$ corresponding to the quoted confidence level is calculated 
by dividing the distribution percentage point probability by the number
of trials, in this case the number of RGS spectral bins ($3400$), under the
approximation that they are independent (this is largely valid for
high-resolution cameras) (Svoboda et al. 2010).
There is no obvious identification for this feature. The nominal
closest transition
according to the ATOMDB database would be a resonant transition between
the ground state and the 
1s4p~$^1$P$_{1}$ electronic configuration of {\sc O vii}. Given the lack of
other detected transitions which could provide information on the velocity
shift of the absorbing gas, as well as the modest statistical significance of
this detections, we consider it as tentative only. We therefore
do not discuss it further. 

On the other hand, several emission lines are detected once a similar
search procedure as described in Guainazzi \& Bianchi (2007) is applied
to the combined time-averaged RGS spectra of \mrk79. Most of
of them are associated to the {\sc N vi} and {\sc O vii} triplets.
The centroid energies of the triplet forbidden components are $431 \pm 3$
and $574 \pm 4\eV$, respectively. The intensity of the triplet components are
reported in Table~\ref{tab1}, together with the values of the
$R$, $L$, and $G$ diagnostic parameters \cite{porquet00,porter07}.
\begin{table}
\caption{Intensities (in units of $10^{-5}$~cm$^{-2}$~s$^{-1}$) and values of the
$R$, $L$, and $G$ diagnostic parameters for the {\sc N vi} and the {\sc O vii}
triplets in the time-averaged \mrk79\ RGS spectra}
\label{tab1}
\begin{tabular}{lcccccc} \hline 
Species & I(f) & I(i) & I(r) & R & L & G \\ \hline
NVI & $2.5\pm 1.5$ & $<2.5$ & $1.4 \pm 1.3$ & $>1.0$ & $<3.8$ & $<4.6$ \\
OVII & $4.9 \pm 1.4$ & $<1.8$ & $<0.7$ & $>2.1$ & ... & ... \\ \hline 
\end{tabular}
\end{table}
The {\sc Ovii} H$_{\beta}$ is also detected: $E_c=671^{+2}_{-5}$~eV,
$I = 1.7 \pm 0.8\times 10^{-5}$~cm$^{-1}$~s$^{-2}$, while the
detection of the O{\sc viii} Ly-$\alpha$ is only marginal ({\it i.e.} lower
than the 90\% confidence level).

The values of the $R$ and $G$ parameters, as well as the high intensity ratio
between the forbidden component of the {\sc O vii} triplet and the {\sc O viii} Ly-$\alpha$
($4.4 \pm 3.3$; Guainazzi \et 2009) indicates photoionisation as the most likely
source of ionisation for the gas responsible for the emission features.
The properties of the photoionised plasma are, however, only poorly constrained:
$n_e<10^{10}$~cm$^{-3}$, $T<10^5\K$, $N_H>10^{20}\pscm$.
These values are consistent with those typically measured from large-scale
photoionised plasmas in obscured AGN (Kinkhabwala \et 2002; Sako \et 2000;
Armentrout \et 2007; Awaki \et 2008).

\section{The 3--12~\keV\ band}
\label{sect:higheband}

\subsection{The existence of a broad line}

Figure~\ref{fig:pofit} shows excess residuals in XMM6 and SUZ1 redward of $6.4\keV$
that could be described by relativistically broadened iron emission.  When fitting broadband 
continuum models such as {\tt reflionx} (Ross \& Fabian 2005) to the $0.3-12\keV$ range,
the fit will be driven by the soft excess since the
statistics are highest below $2\keV$.  Thus the reflection and blurring model parameters 
depend on the interpretation for the soft excess.  Here, only the spectrum above $3\keV$ 
is examined for the existence of broadened iron emission.  Furthermore, only SUZ1 and XMM6
are considered as residuals that can be described with a broad Gaussian profile (Section~\ref{sect:ssf})
are clearly seen at those epochs. This enhances the potential to 
distinguish the power law continuum and reflection component.

\begin{figure}
\rotatebox{270}
{\scalebox{0.32}{\includegraphics{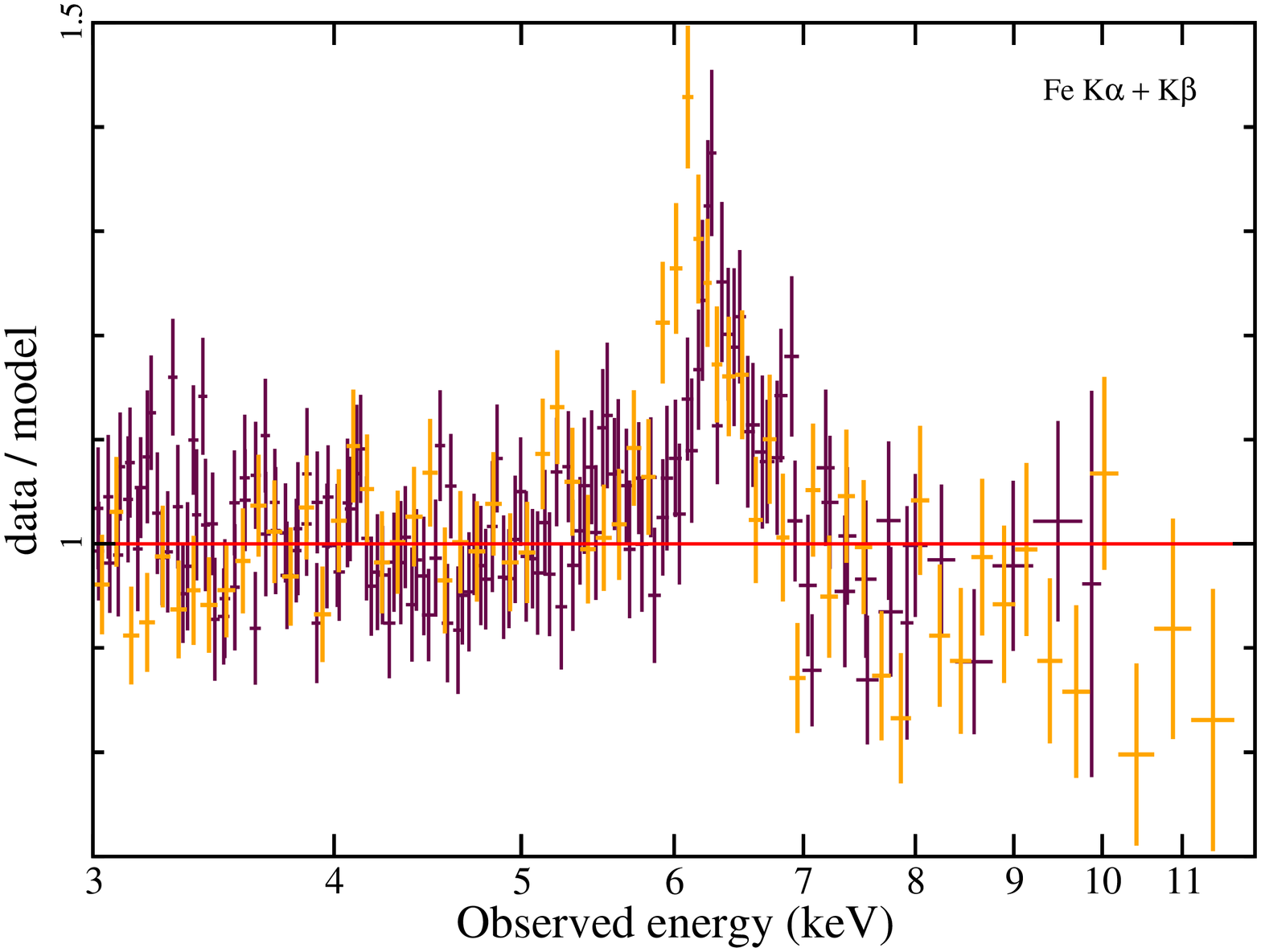}}
\scalebox{0.32}{\includegraphics{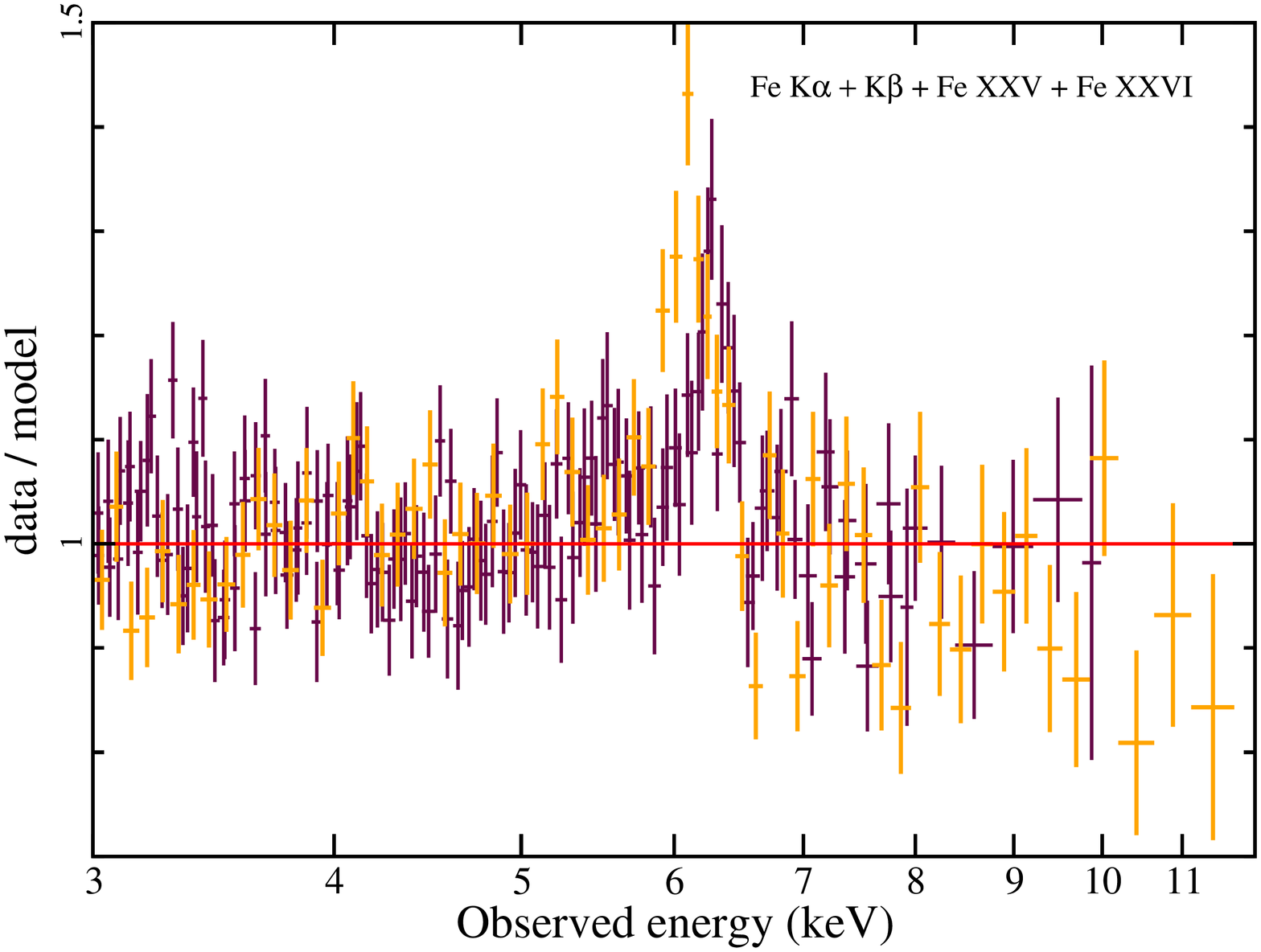}}
\scalebox{0.32}{\includegraphics{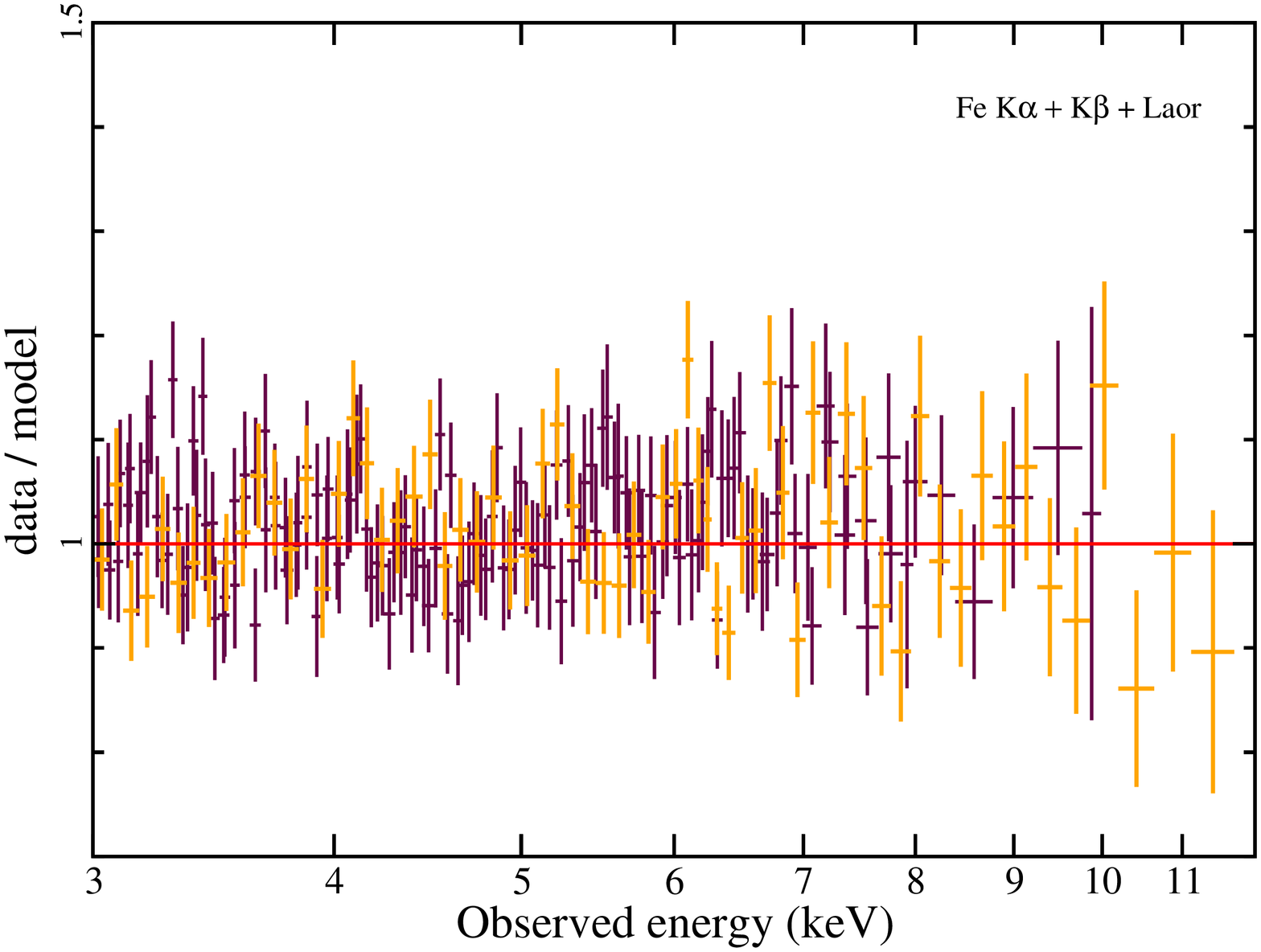}}}
\caption{The residuals (data/model) remaining from fitting the XMM6 (orange) and 
SUZ1 (maroon) spectra
between $3-12\keV$ with a {\tt pexrav} continuum and line models as indicated in
each panel (see text for details).
}
\label{fig:fekab}
\end{figure}

Initially the power law component is replaced with {\tt pexrav} (Magdziarz \& Zdziarski 1995) to
describe the reflection of an exponential cut-off power law from a neutral disc.  The model
does not include an emission line, which is inserted separately.  According to the best-fit
equivalent width ($EW$) measurements of the narrow \feka\ line at each epoch, the reflection fraction
($R = EW/180\eV$; George \& Fabian 1991) is set to 0.58 and 0.89 for SUZ1 and XMM6, respectively.
In addition, two narrow Gaussian profiles representing \feka\ and \fekb\ are included at fixed
energies and relative normalisations (\fekb\ = 0.17\feka).  
The model is statistically acceptable ($\chidof=0.950/2007$)
and fits the narrow core seen in Figure~\ref{fig:pofit} (upper panel), but leaves broad residuals at both epochs 
(top panel Figure~\ref{fig:fekab}).
Two additional Gaussian profiles corresponding to \fexxv\ and \fexxvi\ are included, but
do not improve the fit or residuals significantly (middle panel Figure~\ref{fig:fekab}).
In addition, the flux of \fexxvi\ is consistent with zero in XMM6. 

Replacing the two high-ionisation lines with a single Laor profile (Laor 1991) improves the residuals
significantly (lower panel Figure~\ref{fig:fekab}).  The number of free parameters are the
same as in the model with high-ionisation lines since only the energy and normalisation are
free to vary.  The best-fit energies are $6.60\pm0.08\keV$ (SUZ1) and $6.35\pm0.04\keV$ 
(XMM6).  

In a second fitting attempt with the Laor model, the disc inclination ($i$) and inner disc 
radius ($R_{in}$) are 
allowed to vary, and the emissivity index ($\alpha$) remains fixed.  
This was done in order to estimate some of the parameters for when more complicated
models are attempted (see Section~\ref{sect:multifit}).  
The derived fit parameters are shown in Table~\ref{tab:laor}.  The differing line energies at
each epoch could be suggestive of a change in disc ionisation.  These preliminary fits
argue against a maximally rotating black hole in \mrk79\ since $R_{in}>1.2\rg$ (where $1\rg=GM/c^{2}$). 

\begin{table}
\begin{center}
\caption{Parameter values for relativistic broad line (Laor profile) fits to SUZ1 and XMM6.
Values that are linked between epochs appear in only one column.  Flux units are
\ergpscmps.  The $f$ superscript identifies fixed parameters.
}
\begin{tabular}{ccc}                
\hline
Parameter   &  SUZ1  & XMM6 \\
\hline
$E_{rest}$ (\keV) & $6.77\pm0.12$ & $6.47\pm0.10$  \\
$\alpha$ & $3^{f}$  &  \\
$R_{in}$ ($\rg$) & $6.23^{+1.94}_{-1.12}$ &  \\
$R_{out}$ ($\rg$) & $400^{f}$  &  \\
$i$ ($\deg$) & $21^{+4}_{-3}$ &  \\
log$F_{0.5-10\keV}$ & $-12.63^{+0.08}_{-0.10}$ & $-12.52^{+0.05}_{-0.06}$  \\
\hline
\label{tab:laor}
\end{tabular}
\end{center}
\end{table}

\begin{figure}
\rotatebox{270}
{\scalebox{0.32}{\includegraphics{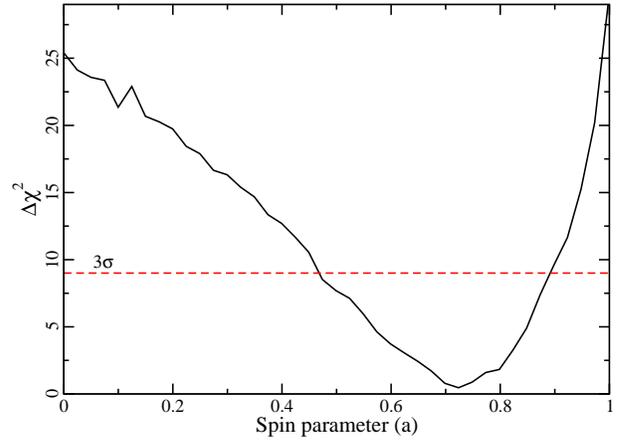}}}
\caption{The quality of the fit is tested against the black hole spin parameter
showing that extreme (high and low) spin values are discriminated against.  The
minimum $\Delta\chi^2$ is at $a\approx0.7$.  The red dashed line marks the $3\sigma$
level.
}
\label{fig:spin}
\end{figure}

\begin{figure}
\rotatebox{270}
{\scalebox{0.32}{\includegraphics{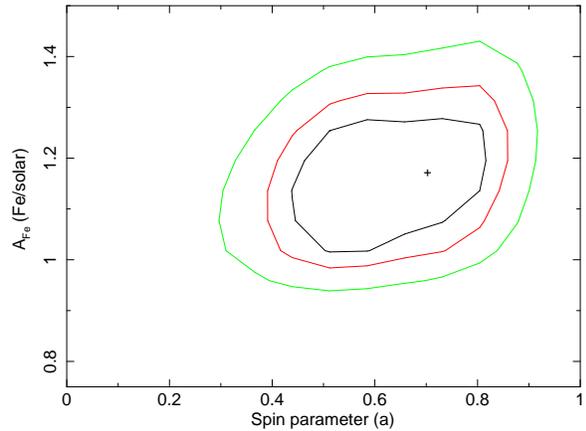}}}
\caption{The $\delchi = 2.30, 4.61, 9.21$ contours between iron abundance and
black hole spin.
}
\label{fig:con}
\end{figure}

Alternatively the $3-12\keV$ spectra are fitted using a power law for the intrinsic continuum
and {\tt reflionx} for the reflection component.  
Blurring of the spectrum,
which is thought to originate close to the black hole, is accomplished with {\tt kerrconv}
(Brenneman \& Reynolds 2006).
In this case, the blurring and reflection parameters: 
inclination ($i$), iron abundance ($A_{Fe}$), black hole spin ($a$), outer disc radius
($R_{out}$) are not expected to vary significantly between the two observations hence are linked.
The power law continuum ($\Gamma$ and normalisation) as well as the ionisation parameter
and normalisation of the reflector are permitted to vary.  The quality of fit is equivalent
to the Laor model ($\chidof=0.836/2005$), but provides a more realistic interpretation for the data.
The measured spin parameter is $a=0.7\pm0.1$.  
In Figure~\ref{fig:spin} variations in the quality of $3-12\keV$ band fit associated with the spin
parameter are plotted for the model described.  Moderately high spin values ($a\sim 0.40-0.85$) 
are favoured. 
We also examine the influence of iron abundance on the spin parameter.  In Figure~\ref{fig:con}
the uncertainties in iron abundance and spin value are plotted together showing that they
influence each other rather modestly.

\subsection{Consideration of absorption models }
\begin{figure}
\rotatebox{270}
{\scalebox{0.32}{\includegraphics{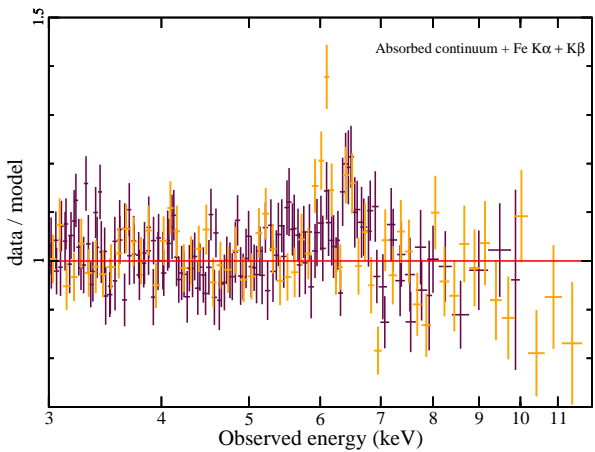}}}
\caption{The residuals (data/model) remaining from fitting the XMM6 (orange) and
SUZ1 (maroon) spectra between $3-12\keV$ with ionised partial covering continuum and
line models as indicated (see text for details).
}
\label{fig:abslin}
\end{figure}

Miller \et (2007, 2009) have described AGN X-ray spectra as being shaped by a composition of
ionised absorbers partially covering the primary source.  In doing so, their model  diminishes the necessity of a very broad 
relativistic iron lines in some cases.  The addition of either one or two ionised absorbers
to modify the continuum of \mrk79\ produces an equally good fit as the models described above.
The appearance of the emission feature is reduced, but not dismissed (Figure~\ref{fig:abslin}).
The addition of a Gaussian or Laor profile improves the residuals.  In terms of a Gaussian profile
the best-fit parameters for the SUZ1 and XMM6 data are respectively: $E=6.54\pm0.08$ and $6.34\pm0.04\keV$;
$\sigma=171^{+60}_{-81}$ and $199^{+43}_{-30}\eV$; and flux $F=1.23\pm0.04$ and $2.93\pm0.03\times10^{-13}\ergpscmps$.

\section{A multi-epoch self-consistent model for \mrk79}
\label{sect:multifit}

The above analysis demonstrates that the spectra of \mrk79\ are rather complex and a 
composition of multiple effects.  In this section we will treat each phenomenon with
more physical models while attempting to fit the data from all epochs together.  The
broadband, multi-epoch analysis is driven by the findings in the previous sections.

\subsection{The baseline model: ionised disc reflection}
\label{sect:spinfit}
The positive residuals above a power law continuum between $6-7\keV$ seen in the low-flux
states can be attributed to broadened \feka\ emission from the accretion disc near the black hole.
This is typically the strongest signature of reflection in AGN X-ray spectra (e.g. Miller 2007).
Partial covering models can also describe the continuum in the low-flux states, but do not eliminate
the residuals, and consequently require a second component.
 
Illumination of a modestly ionised accretion disc by a power law continuum source will also
generate fluorescence of low-ionisation species (e.g. \fela; Fabian \et 2009; Zoghbi \et 2010) below $\sim2\keV$ 
(Ross \& Fabian 2005; Ballantyne \et 2001).  
Once blurred for relativistic effects close to the black hole it has been shown that the low-energy 
reflection component
can mimic the black body-like soft excess seen in most type~I AGN (e.g. Crummy \et 2006).

As a starting point for fitting all the \mrk79\ spectra, the power law plus {\tt reflionx} model used to fit the 
$3-12\keV$ band during
the low-flux states (i.e. XMM6 and SUZ1, see Section~\ref{sect:higheband}) is applied to the broadband
spectrum at all flux states. 
Even though a prominent broad line is not obvious in the high-flux state spectra,
expectations are that a reflection component is still present (i.e. the power law source still illuminates
the accretion disc),
but that its flux is diminished relative to the continuum.
As with the analysis in Section~\ref{sect:higheband}, {\tt kerrconv} is used to blur the reflection component.
One advantage of fitting multi-epoch data simultaneously is that parameters not expected
to vary on the time scales probed can be linked.  In this case, the reflector and blurring parameters: 
inclination ($i$), iron abundance ($A_{Fe}$), black hole spin ($a$), outer disc radius
($R_{out}$) are not expected to vary on time scales of 8-years hence are linked between all epochs.

At all epochs, the inner disc radius ($R_{in}$) is assumed to be located at the innermost stable circular orbit (ISCO), 
hence fixed at $R_{in}=1.0\risco$.  Initial fits always place the outer disc radius at large distances so for simplicity
$R_{out}$ is fixed at $100.0\risco$.  
The emissivity profile of the disc is a power law in radius ($J(r)\propto r^{-\alpha}$) where $\alpha=3$
is commensurate with a simple lamp post model.  The index is left free to vary at each epoch.
Including a different index for the inner and outer portion of the disc 
(i.e. a broken power law emissivity profile) was initially considered, but deemed
unnecessary.  The two indices were usually comparable and/or the break radius separating the inner
and outer disc regions was very large (e.g.$\gs150\rg$).
The spin parameter is fixed to $a=0.7$, the value calculated when fitting only the low-flux
states (Section~\ref{sect:higheband}). Likewise the iron abundance is set to $1.2\times$. 
This was primarily done to keep from generating arbitrarily high spin values that 
would artificial smooth the spectrum at low-energies (i.e. the soft excess) where statistics are best.
The ionisation parameter of the disc ($\xi=L_{x}/nr^{2}$, where n is the hydrogen number density),
power law continuum ($\Gamma$ and normalisation), 
and normalisation of the reflector are permitted to vary at each epoch.

The blurred reflection and power law serves as a reasonable baseline model 
for all the spectra ($\chidof=1.247/7425$).  It results in similar
residuals as found with a black body and power law, specifically
a narrow emission feature at \feka\ energies and sharp wiggles at lower energies.
The residuals in the \feka\ complex are treated by adding two narrow Gaussian profiles to mimic emission. 
The energy of one Gaussian profile is fixed at $7.058\keV$ 
representing emission from \fekb. 
The second Gaussian profile, likely due to distant \feka\ emission, is free to vary
in energy, width and normalisation.  The width of the \fekb\ profile is linked to the free line.
In addition, the flux of the \fekb\ line is constrained 
to be $0.17$ times that of \feka.  
Narrow emission lines from highly ionised species of iron (\fexxv\ and \fexxvi) were
considered unnecessary in Section~\ref{sect:higheband} and consequently not modelled here. However the 
residuals of each spectrum are examined for possible features.
The normalisation of
the \feka\ line is linked between all the spectra as there was no indication of it being 
variable from one epoch to another (see Section~\ref{sect:ssf}).  In total, the addition
of two Gaussian profiles account for an extra three parameters in the fit.

\subsection{The photoionised emitter and warm absorber}
The RGS analysis in Section~\ref{sect:rgs} reveals possible absorption, but predominately emission, 
from an ionised plasma.  The dominance of the forbidden component in the \ovii\ triplet and high 
\ovii (f)/\oviii\ Ly$\alpha$ point to photoionisation as the primary source of ionisation.  
The {\tt XSTAR} (Kallman \et 1996) analytical models
{\tt warmabs} and {\tt photemis} are used to model the warm absorption and thermal (i.e. recombination and
collisional excitation) emission, respectively.  The ionised plasma is assumed to have a 
density of $10^{4}\pccm$ consistent with the density estimated in the RGS analysis.  Element
abundances were frozen at solar values.
The absorber is assumed to cover the power law and reflection components.

There is significant freedom in how the ionised plasma is modelled at each epoch and different approaches 
are considered here.  For starters, the parameters of the photoionised emitter are linked between
all epochs.  One possibility to handle the absorber is to link the column density at all epochs and allow
only the ionisation to vary. This is a considerable improvement over the baseline model
discussed above ($\chidof=1.028/7412$).  Statistically, a better quality of fit is achieved when
the ionisation of the absorber is linked and the column density is free to vary at each epoch
($\delchi=-84$ for the same number of variables).  However, in both cases sharp residuals
remain at low energies in some spectra.  Allowing both the column density and ionisation to vary
at each epoch is a significant improvement over fits where only one of the parameters is free
to vary ($\delchi=78$ for 7 additional free parameters).  In addition the residuals at low energies
are also improved ($\chidof=1.006/7406$).  Allowing the photoionised emitter to vary in
ionisation and normalisation at each epoch is only a marginal improvement to the fit
($\chidof=0.999/7394$).  However, it is physically feasible for the emitter to vary on
such time scales and since the absorber is permitted to vary this is a consistent approach.

\begin{figure}
\rotatebox{270}
{\scalebox{0.32}{\includegraphics{fig12a.ps}}
\scalebox{0.32}{\includegraphics{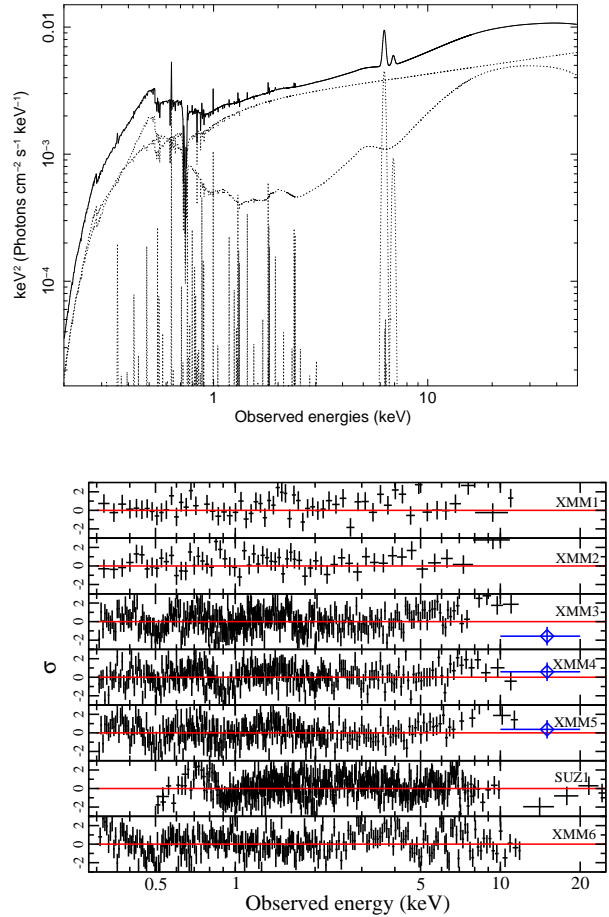}}}
\caption{Upper panel: The model used to fit the spectrum of XMM1.  XMM1 represents
and intermediate flux state for \mrk79.  The model is 
similar for all the spectra so only one spectrum is shown for clarity.
Lower panel: Spectral residuals (in sigma) at each observation resulting from the simultaneous multi-epoch fit described
in the text and Table~\ref{tab:fits}.
The blue diamonds in the residuals are based on simultaneous $10-20\keV$ 
flux measured with \rxte. They are shown to compare the model predicted flux above $10\keV$
with data, but were not part of the fit.
}
\label{fig:reffit}
\end{figure}

The described model produces an acceptable fit to the multi-epoch spectra of \mrk79 ($\chidof=0.999/7394$).
In Figure~\ref{fig:reffit}, the model used to fit the intermediate-flux state of \mrk79\
(XMM1) is shown 
as well as the residuals resulting from the fit for each spectrum.  XMM3,
XMM4, and XMM5 benefit from simultaneous \rxte\ observations in which a $10-20\keV$
flux could be measured.  These points are included on 
Figure~\ref{fig:reffit} (blue diamonds) to show the very good agreement between the model predicted 
flux and the measured flux, but they were not fitted.  The model parameters at each epoch are
presented in Table~\ref{tab:fits}.
We reiterate that parameter uncertainties will be underestimated in this section.  
Given the complexity of the model,
when calculating uncertainties for a given parameter the other variables were held fixed.
Consequently, the variability in a given parameter from one epoch to the next is not as significant
as it appears, and examining general trends in the variations is perhaps more valuable.
\begin{table*}
\caption{Best-fit model parameters for the multi-epoch modelling of \mrk79.  The model component and model parameter is listed in Columns 1 and 2, respectively.
Each subsequent column refers to a specific epoch.  The inner ($R_{in}$) and outer ($R_{out}$) disc radius are given in units of the radius of marginal stability ($r_{ms}$), where we have further assumed that $r_{ms}=\risco$.  
The log of the ionisation parameter is shown for both ionised plasmas and the warm absorber column density is in units of log(N/$10^{22}\pscm$).
Values that are linked between epochs appear in only one column.  
The $f$ superscript identifies fixed parameters.
Fluxes are corrected for Galactic absorption and are in units of $\ergpscmps$.}
\centering
\scalebox{0.8}{
\begin{tabular}{ccccccccc}                
\hline
(1) & (2) & (3) & (4) & (5) & (6) & (7) & (8) & (9) \\
Model Component &  Model Parameter  &  XMM1 & XMM2 & XMM3 & XMM4 & XMM5 & SUZ1 & XMM6 \\
\hline
Power law & $\Gamma$ & $1.76\pm0.01$ & $1.95\pm0.01$ & $2.00\pm0.01$ & $1.81\pm0.01$ & $1.86\pm0.01$ & $1.61\pm0.01$ & $1.49\pm0.01$ \\
          &log $F_{0.5-10 keV}$ & $-10.835\pm0.007$ & $-10.498\pm0.006$ & $-10.440\pm0.002$ & $-10.526\pm0.002$ & $-10.593\pm0.003$ & $-10.712\pm0.002$ & $-11.164\pm0.003$ \\
\hline
Warm Absorber & log $n_H$ & $-0.68\pm0.03$ & $-0.83\pm0.03$ & $-1.11\pm0.02$ & $-1.10\pm0.02$ & $-1.10\pm0.02$ & $-0.86\pm0.02$ & $-1.08\pm0.06$ \\ 
              & log$\xi$ & $1.03^{+0.04}_{-0.07}$ & $1.94^{+0.26}_{-0.08}$ & $0.98\pm0.04$  & $0.81^{+0.09}_{-0.04}$ & $1.00\pm0.04$  & $0.93^{+0.08}_{-0.04}$ & $1.73\pm0.03$   \\
\hline
Photoemitter  & log$\xi$ & $2.09^{+1.22}_{-0.62}$ & $1.03^{+1.00}_{-1.03}$ & $2.07^{+0.25}_{-0.09}$  & $2.13^{+0.41}_{-0.21}$ & $2.08^{+0.25}_{-0.17}$ & $2.08^{+0.68}_{-0.21}$  & $1.38^{+0.05}_{-0.03}$  \\
              & log $F_{0.5-10 keV}$ & $-13.03^{+0.26}_{-0.71}$ & $-13.00^{+0.37}_{-1.04}$  & $-12.97^{+0.21}_{-0.41}$  & $-13.16^{+0.25}_{-0.64}$ & $-13.09^{+0.23}_{-0.50}$ & $-13.43^{+0.28}_{-1.00}$ & $-12.80\pm0.03$    \\
\hline
Narrow Lines  & $E_{FeK\alpha}$ ($\keV$) & $6.40\pm0.01$ &                         &                &                &                &                &                          \\
              & $E_{FeK\beta}$ ($\keV$) & $7.058^{f}$       &                         &                &                &                &                &                           \\
              & $\sigma$ ($\eV$)  & $99^{+15}_{-9}$ &                      &                &                &                &                &                           \\
              & log $F_{FeK\alpha}$ & $-12.54\pm0.03$ &                  &                &                &                &                &                           \\
\hline
Blurring      & $\alpha$    & $4.62^{+0.31}_{-0.27}$ & $4.16^{+0.51}_{-0.40}$ & $4.83\pm0.11$          & $3.78\pm0.13$          & $5.53\pm0.13$          & $3.27^{+0.19}_{-0.17}$ & $4.13\pm0.10$          \\
              & $a$         & $0.7^{f}$ &                       &                        &                        &                        &                        &                        \\
              & $R_{in}$   & $1.0^{f}$ &               &                        &                        &                     &                        &                     \\
              & $R_{out}$   & $100^{f}$                  &                        &                        &                        &                        &                        &                         \\
              & $i$ ($\deg$)  & $24\pm1$               &                        &                        &                        &                        &                        &                         \\
\hline
Reflection    & $\xi$ ($\erg\cmps$)       & $179\pm5$        & $57\pm4$               & $131\pm3$      & $130\pm2$      & $145\pm2$        & $177\pm6$      & $60\pm1$          \\
              & $A_{Fe}$ (Fe/solar)          & $1.2^{f}$              &                        &                &                &                  &                &                   \\
          &log $F_{0.5-10 keV}$ & $-11.126\pm0.014$    & $-11.174\pm0.029$        & $-10.804\pm0.005$      & $-11.154\pm0.008$      & $-10.892\pm0.005$      & $-11.551\pm0.017$        & $-11.439\pm0.006$       \\
\hline
Total flux   &log $F_{0.5-10 keV}$ & $-10.733\pm0.005$    & $-10.445\pm0.005$        & $-10.354\pm0.001$      & $-10.470\pm0.002$      & $-10.492\pm0.002$      & $-10.674\pm0.002$        & $-10.962\pm0.002$       \\
             &log $F_{2-10 keV}$ & $-10.929\pm0.005$    & $-10.701\pm0.005$        & $-10.631\pm0.001$      & $-10.687\pm0.002$      & $-10.736\pm0.002$      & $-10.834\pm0.002$        & $-11.083\pm0.002$       \\
\hline
\label{tab:fits}
\end{tabular}
}
\end{table*}

A complexity that arises in accepting the parameter uncertainties at face value is that the
model predicts an apparent correlation between power law flux
and photon index, in the sense that the intrinsic spectra are harder in the low-flux states.
This prediction is different from the general interpretation of reflection or absorption models
where the intrinsic slope is constant, but the apparent slope changes due to varying contribution
of the separate components.  We reconsider the error estimation for $\Gamma$ in the
XMM6 and SUZ1 spectra individually.  While maintaining all multi-epoch determined parameters
fixed (e.g. inclination, abundances) and varying the remaining parameters, we estimate the more
likely uncertainty in $\Gamma$ at each low-flux state.  In both cases, $\Delta\Gamma\approx 0.1$
at a $3\sigma$ level.  While this alleviates the concern for significant intrinsic power law
shape changes during SUZ1, it is still problematic during XMM6.
Intrinsic spectral changes in the power law component are predicted 
if the primary emission originates from a jet or Comptonisation, but the large changes
from XMM6 to other epochs are difficult to reconcile.  

\section{Discussion} 
\subsection{The photoionised plasma}
\begin{figure}
\rotatebox{270}
{\scalebox{0.32}{\includegraphics{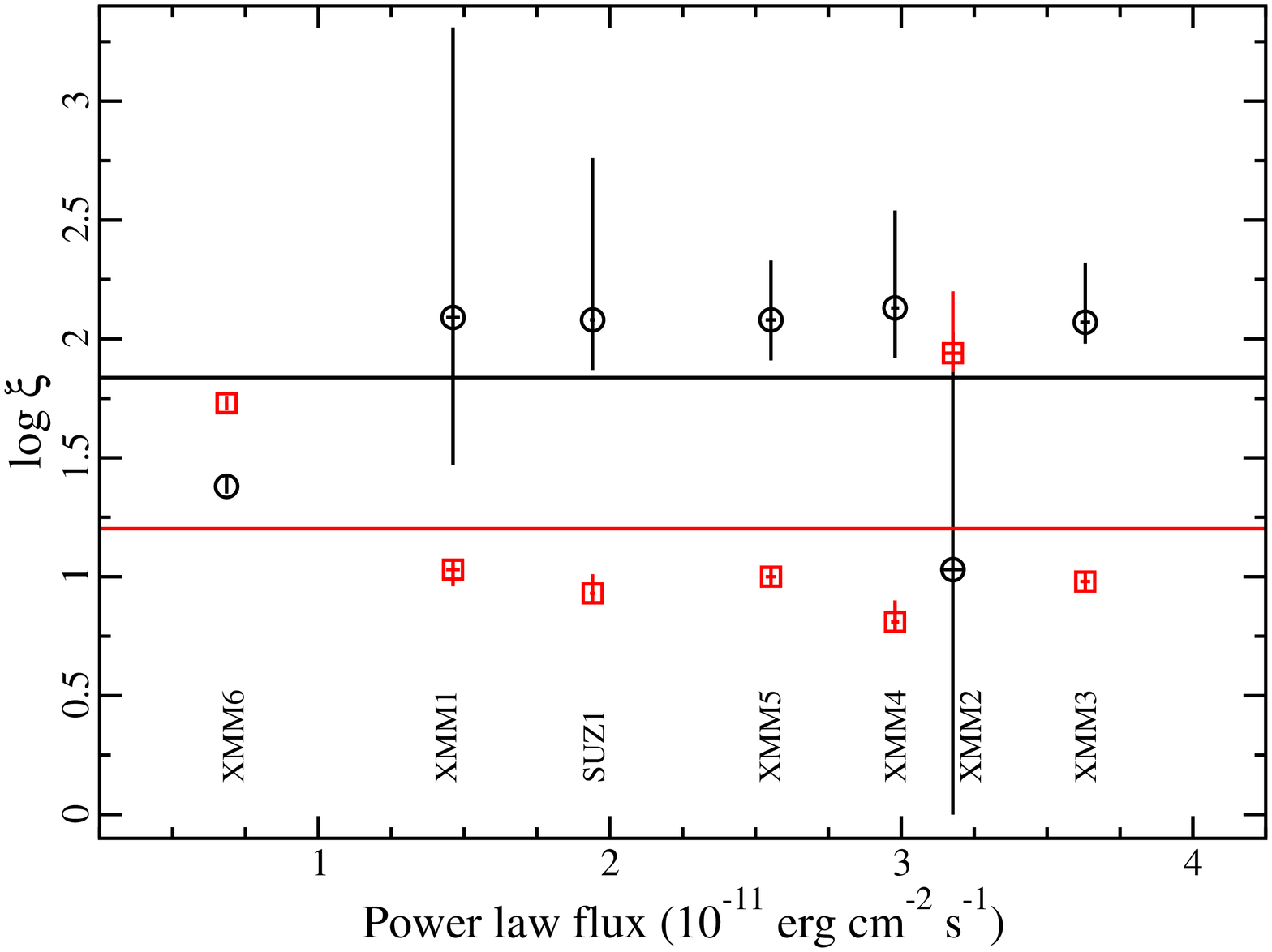}}
\scalebox{0.32}{\includegraphics{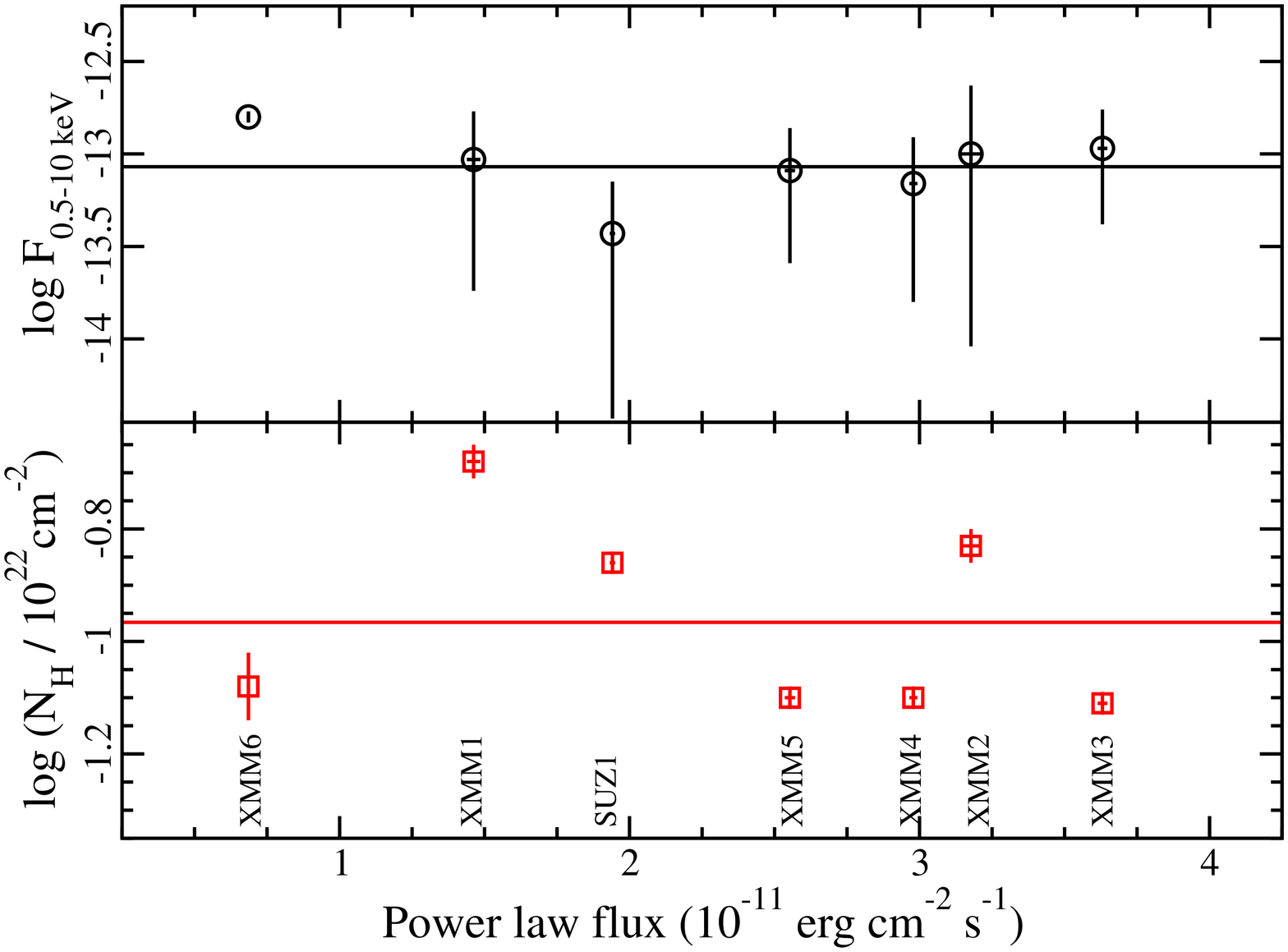}}}
\caption{Upper panel: The logarithm of the ionisation parameter of the photoionised emitter
(black circles) and absorber (red squares) is plotted against the
flux of the power law component.  The average ionisation parameters for each component 
are also shown (straight lines).  
Lower panel: The $0.5-10\keV$ flux of the emitter (black circles) and column density 
of the absorber (red squares) is shown against power law flux.
Uncertainties are plotted on all quantities, but may be smaller than the data point
itself.
}
\label{fig:plasma}
\end{figure}

The state of the photoionised emitter and warm absorber shows no correlation with the 
flux of the power law emitter (Figure~\ref{fig:plasma}).  This alone does not
dismiss a connection between the power law and plasma since the sparse
data makes it impossible to track delays caused by light travel 
time effects 
and recombination time scales.  
Of course, the light travel time delays would be zero for the line-of-sight
absorber.

On average the emitter exhibits a higher ionisation parameter
than the absorber indicating that they are different structures.  If both are
ionised by the AGN and have similar densities, then the absorber is the more distant of the two.
Using the ionisation parameter to estimate the distance ($r=(L_{x}/n\xi)^{1/2}$)
between the AGN and absorber places the absorber at $\sim2\times10^{19}\cm$ ($6.5\pc$).
The emitter would be approximately half as far from the AGN.
Both are too distant from the AGN to detect correlations with AGN luminosity from our
data.

The ionisation parameter of the emitter is comparable at all epochs except for
the lowest flux state (XMM6) where the level of ionisation decreases.  It is not
possible to determine if this change is driven by some prolonged decrease in the
continuum flux of the AGN at some point in the past.
The emitter shows no significant
changes in flux level.

The column density of the near-neutral warm absorber is on the order of $10^{21}\pscm$, 
comparable to hydrogen column densities seen in our own Galaxy and could
be attributed to a similar structure in \mrk79.
The column density does exhibit some variability ($\sim20$~per cent).
Such variations are not very common in type~I AGN, but have been reported previously
(e.g. Gallo \et 2007).

\subsection{Bulk motion of the primary source and the role of light bending in \mrk79}
Objects like IRAS~13224--3809 and 1H~0707-495, where light bending is attributed with
a substantial role,  often exhibit high values for the emissivity index (e.g. $\alpha>6$)
(e.g. Ponti \et 2010; Zoghbi \et 2010), indicating X-ray emission originating from
a centrally compact region.  In \mrk79\
the best-fit model was achieved when disc emission over a large area was allowed.
This is evident by the relatively shallow emissivity profile measured for the disc.
The emissivity index in \mrk79\ is typically between $\alpha\approx3-5$ 
and this may even be slightly overestimated since, in general, limb-brightening effects
are not considered in our blurring model (Svoboda \et 2009).  The emissivity in
\mrk79\ is commensurate with a simple lamp post model ($\alpha=3$).  

\begin{figure}
\rotatebox{270}
{\scalebox{0.32}{\includegraphics{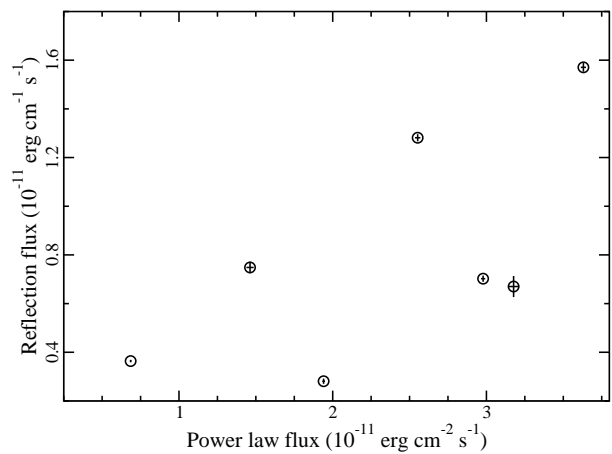}}}
\caption{The $0.5-10\keV$ unabsorbed flux of the reflection and
power law components are plotted.  Variability of similar amplitude
is present in both.
}
\label{fig:refpo}
\end{figure}

In the light bending model the primary component is normally perceived as an isotropic and
relatively constant emitter.  The observed rapid variability, as described by Miniutti \& Fabian (2004),
is driven by changes in the distance of the primary emitter 
from the black hole/accretion disc.
The model makes specific predictions on the temporal behaviour of
the reflection and power law component.  For example, variations in the
two components are correlated when the illuminating source is within a
few $\rg$ of the black hole.  At intermediate distances ($\sim10\rg$)
the reflection component remains rather constant while the power law can
vary significantly.
On long-time scales, as observed in this work, the power law and reflection components will
be largely correlated assuming that the illuminating component is emitting isotropically.
In addition, light bending does not exclude the possibility of long term intrinsic
changes in the flux of the primary emitter.  Given that these data span nearly eight
years, long-term intrinsic changes could take place as well.
In Figure~\ref{fig:refpo} average flux values for each observation are plotted for
the reflection and power law components.  The figure points to
variations of similar amplitude in both components.
There is a common trend between the fluxes of the power law and reflection,
but not a one-to-one correlation.

More importantly, the relative contribution of the reflection
component in \mrk79\ is small.  
We estimated the reflection fraction ($R$) by comparing
{\tt pexrav} with simulations of the spectral models above $10\keV$ at each epoch.  
The power law slope, abundances and inclination are fixed to the values derived with
the fits (Table~\ref{tab:fits}).  The cutoff energy is fixed at $350\keV$, the value 
used by {\tt reflionx}.
At all epochs the reflection
fraction was between
$R \approx 0.24-0.73$, with the highest $R$ value occurring for XMM6.  
Notably, $R$ is variable from epoch-to-epoch and is consistently less than one.
While reflection features are present in \mrk79, light bending does not appear to 
be a significant factor.

The variations in $R$ also advocate a non-standard geometry for the illuminating source.
If the source were
simply varying its luminosity to produce the observed variability,
there should exist a good correlation between the reflection and power law flux along with 
a constant reflection fraction.  This is not the case.  Therefore the standard picture in
which the primary source, wherever it is located, only varying in luminosity with
its other properties remaining unchanged is challenged.

One can imagine a situation in which a compact primary source is located at a few $\rg$, hence
subject to general relativistic effects, but is moving away at mildly relativistic speeds
(e.g. Reynolds \& Fabian 1997; Beloborodov 1999).
Beloborodov proposed such a model to explain the low reflection fraction in the black hole binary Cyg~X-1.
Beloborodov described the nature of the primary source as an active corona atop the accretion disc.
The plasma in the active regions is heated by magnetic flares, and it can be ejected at
mildly relativistic velocities if the blob is dominated by $e^{\pm}$ pairs.
As a consequence of beaming the reflection fraction is reduced even in the presence of light bending
effects.  
From equation (3) of Beloborodov (1999),
$R = (1- \beta/2)(1-\beta cos i)^3 / (1+\beta)^2$,
the bulk velocity of the illuminating blob ($\beta=v/c$) is related to $R$, where $i$ is the inclination
of the system.
We determine $\beta$ at each epoch from the estimated reflection fraction.
The maximum velocity will occur at the lowest $R$ ($\approx 0.24$) and in this case is $\beta\approx0.32$.
As the source is radio-quiet one would not expect these ejections to escape.  
For the estimated $\beta$,
the escape radius from the $5.24\times10^7\Msun$ black hole in \mrk79 is $\approx20\rg$.
For higher
values of $R$, one can allow for smaller $\beta$, which would result in sources within an even larger distance.
Light bending could have a diminished role in \mrk79\ if the source is dominated by beaming.
The fact that \mrk79\ tends to be X-ray strong for its UV luminosity (Figure~\ref{fig:uv}) is
consistent with the X-rays being beamed.

As a consistency check we used equation (10) of Beloborodov, 
$\Gamma \approx 2((1-\beta^2)^{-1/2}(1+\beta))^{-0.3}$, to compare the predicted photon index with
our measured values (Table~\ref{tab:fits}).  There is very good agreement between the measured and predicted
$\Gamma$ (within a few per cent)
for the high-flux observations (XMM1--XMM5).  
For the low-flux observations (SUZ1 and XMM6) the difference is $10-20$ 
per cent, and alone, the  
$\Gamma(\beta)$ relation for this active corona model 
cannot explain the low-state spectral hardening.  As discussed in Section~6, this is more problematic
for XMM6.
We note that SUZ1 and XMM6  were the longest observation, and did exhibit some spectral variability
within the observation (Figure~\ref{fig:fvar}) 
that would affect fitting of the average spectrum.

From long-term optical and X-ray monitoring of \mrk79, Breedt \et (2009)
find that the optical ($V$-band) and X-ray are highly correlated with a delay of $0^{+2}_{-4}$~days.
Breedt \et find that the optical variations could still be produced by thermal
reprocessing of variable X-rays if the height of the illuminating X-ray source
is on the order of $21.5\rg$.  At $\sim20\rg$ the light travel-time delay between
the illuminating X-ray source and the disc is $\tau=20\rg/c=5000\s$.
Consequently, a delay of $\tau\approx5-10\ks$ between the power law and reflection component
is possibly detectable.  Despite all the high-quality data in this study, 
the XMM1--XMM5 observations last $\ls20\ks$
and the \suzaku\ data include gaps in the light curve every $\sim5.6\ks$ due
to its low-Earth orbit rendering them unsuitable to search for a $\gs5\ks$ lag.  
Only XMM6 is sufficiently long enough to search for such a lag, 
and interestingly the cross correlation between the  $0.2-0.8\keV$
(reflection component) and $1.5-2\keV$ (power law component) band
does suggest a shift in the peak to negative times (i.e. high-energy leads
low-energy changes).  However, the XMM6 light curves suffers from the 
fact that it drops smoothly with time and lacks distinct features to build a
correlation on (Figure~\ref{fig:lc}).  
Any detection of a lag with this light curve is judged questionable.

Coincidently, Breedt et al. (2009) find a best-fit disc
inclination of $24\deg$ from modelling the long-term multiwavelength light curve.
We determine a similar inclination $i=24\pm1\deg$ in our independent X-ray spectral 
analysis of the source.

On average, the analysis of \mrk79\ is consistent with the illuminating (power law) source 
being beamed and located at large distances ($\gs20\rg$)
from the disc.

\subsection{Limitations in measuring the black hole spin}
Data quality and spectral models have reached such sophistication that 
the black hole spin parameter can now be measured from the broadened \feka\
line in Galactic black holes
(e.g. Miller \et 2009) and AGNs (e.g. Brenneman \& Reynolds 2006; Miniutti \et 2009;
Schmoll \et 2009).  

For \mrk79, a moderately high spin value is measured ($a = 0.7\pm0.1$) by modelling only
the high-energy spectrum.  Under the assumption that the soft excess in AGNs is
the reflection component it becomes tempting to use the high signal-to-noise data
below $3\keV$ to constrain black hole spin.  In principle this is fine, but
one must acknowledge how complicated this spectral region is.  In addition to uncertainties
in the calibration, there may be unmodelled complexities in the ionised absorber/emitter
and in the reflection spectrum itself.  These complexities could strongly affect conclusions
about spin.

Figures~\ref{fig:spin}
and~\ref{fig:con} also seem to exclude the possibility of a maximally spinning black hole.
In general, it is also difficult to conclusively exclude extremely high spin values since
our models are built under the assumption that the disc is truncated at the ISCO. 
One cannot exclude the possibility that a system includes a maximal spinning black hole, but
that the disc is truncated at a larger radius.
The broad line in \mrk79\ is particularly weak (Figure~\ref{fig:fekab}) and could be consistent
with emission originating from large radii.

\section{Conclusions}

Broadband X-ray spectra ($0.3-25\keV$) of the Seyfert 1.2 galaxy \mrk79\ are analysed
in attempt to examine long term variability.   
The data are obtained with \xmm\ and \suzaku\ at seven different epochs spanning 
nearly eight years, and cover a factor
of about three in broadband flux between low- and high-flux states.  The short time scale
(i.e. within an observation) flux and spectral variations are typically small making
\mrk79\ useful for studying the long-term spectral variability and ``average'' spectrum.  
The main results are as follow.
\begin{itemize}
\item[(1)]
The multi-epoch, broadband continuum is consistent with a primary
power law flux component and disk reflection, both modified by absorption
and emission from an ionized plasma.  The relative weakness of the disk
reflection features in these spectra is consistent with a model for
stellar-mass black holes wherein the power law continuum arises in flares
that are driven vertically from the disk.  Variations in the illuminating
and reflected components do not suggest that light bending dominates the
observed flux trends.

\item[(2)]
Under the assumption that the inner disc is truncated at $\risco$ a moderately
high black hole spin parameter is found ($a=0.7\pm0.1$).  

\item[(3)]
Low-density, photoionised plasma seen predominately in emission is observed and consistent with
arising at large distances from the black hole.  
The measured parameter values are consistent with those typically measured from large-scale
photoionised plasmas in obscured AGN.

\item[(4)]
Simultaneous UV photometry was obtained for 6 of the 7 observations.  The UV are variable
to a lesser degree than the X-rays.  
Measurements of $\alpha_{ox}$ show that \mrk79\ tends to be
in an X-ray bright state compared to its UV luminosity.  This would be consistent with the notion
that the X-rays are beamed.  
\mrk79\ was caught in
an X-ray weak state during only one observation (XMM6) for certain.
In the X-ray weak state the $2-10\keV$ spectrum appears harder and deviates significantly
from a simple power law.

\end{itemize}

The modest intra-observation variability exhibited by \mrk79\ introduces the potential to obtain
a high signal-to-noise average spectrum of the source.  The multi-epoch monitoring 
allows for several model parameters to be constrained with high precision allowing for a more
accurate determination of the variable components.
\mrk79\ does not appear to be a particularly extreme active galaxy, but perhaps it is more representative of  
a typical
AGN, and consequently it is important to scrutinise it.  
{\it Astro-H} and {\it IXO} will have substantially better energy resolution
and hard X-ray
sensitive than current X-ray missions thereby improving our ability to constrain the
warm emitter/absorber and reflection component.


\section*{Acknowledgments}

The \xmm\ project is an ESA Science Mission with instruments
and contributions directly funded by ESA Member States and the
USA (NASA). 
Thanks to Damien Robertson for $\alpha_{ox}$ measurements and
Phil Uttley for the \rxte\ light curve.




\bsp
\label{lastpage}
\end{document}